\documentstyle[aps,epsfig,multicol]{revtex}

\begin{document}

\draft

\title{Eigenfrequencies of two mutually interacting gas bubbles in an acoustic 
field}
\author{Masato Ida}
\address{Satellite Venture Business Laboratory, Gunma University, 
1--5--1 Tenjin-cho, Kiryu-shi, Gunma 376-8515, Japan\\
E-mail : ida@vbl.gunma-u.ac.jp
%\\ TEL : +81-277-30-1126 \quad FAX : +81-277-30-1121
}
\date{\today}

\maketitle

\begin{abstract}
Eigenfrequencies of two mutually interacting gas bubbles in an acoustic field 
are discussed theoretically and numerically. It is shown by a linear theory 
that a bubble interacting with a neighboring bubble has three eigenfrequencies 
that change with the distance between two bubbles, and the sign and magnitude 
of the primary Bjerknes force acting on bubbles also change according to the 
change in the eigenfrequencies.
\end{abstract}

\begin{multicols}{2}

\section*{INTRODUCTION}
It is known that many bubbles levitated in a liquid form a stable structure 
named ``bubble grapes'' when a weak standing sound wave is applied 
\cite{1,2,3,4}. In this stable structure, bubbles or clusters of them maintain 
a distance each other and tend not to collide, contrary to predictions given 
by the classical theory by Bjerknes which allows either only attraction or 
repulsion. The distance between bubbles or clusters in the structure is 
comparable to their sizes. A similar phenomenon can be observed in a strong 
acoustic field \cite{5} in which bubbles form a complex 
filamentary structure varying slowly, whose scale is much smaller than the 
wavelength of the acoustic field.

In order to understand these phenomena, one should know thoroughly how the 
mutual interaction of the bubbles influences their acoustic properties. In 
Ref.~\cite{2}, Zabolotskaya showed theoretically that the effective resonance 
frequencies of two interacting bubbles rise or fall as bubbles approach each 
other and that the variation in the resonance frequencies may sometimes causes 
the change of the 
sign of the secondary Bjerknes force, which is an interaction force acting 
between pulsating bubbles, at a certain distance. In Ref.~\cite{3}, Doinikov 
and Zavtrak introduced a similar result by taking the multiple scattering of 
sound between two bubbles into account. They explain the origin of the change 
in the sign of the force as follows: When two bubbles oscillate in phase, their surfaces move against each other resulting in a stiffening of the 
bubbles. This may increase the effective resonance frequencies (or 
eigenfrequencies) of both bubbles and may sometimes cause the change in the 
sign of the force from attractive to repulsive.

Influences of mutual interaction have also been studied in regard to 
sonoluminescence \cite{7,8}. In Ref.~\cite{7}, Mettin et al. examined 
numerically the sign and magnitude of the secondary Bjerknes force acting 
between two small air bubbles in a strong acoustic field. They determined 
that even when the distance between two bubbles is much larger than their 
equilibrium sizes the magnitude of the force becomes much larger than that 
predicted by a linear theory, and the sign of the force changes near the 
dynamical Brake threshold or around the \textit{nonlinear} resonance frequency. 
In Ref.~\cite{8}, Doinikov showed that a neighboring bubble drastically varies 
the magnitude and sign of the primary Bjerknes force, which an incident sound 
wave exerts on a bubble, in a strong acoustic field.

In this paper, following Zabolotskaya's approach, the influences of the mutual 
interaction on the effective resonance frequencies of 
bubbles are theoretically examined in detail. As was pointed out and 
was mentioned previously, the variations in eigenfrequencies strongly affect 
the primary and secondary Bjerknes forces. This means that full understanding 
of the eigenfrequencies surely must be of help in knowing how the forces are 
changed by mutual interaction. In Sec.~\ref{sec2a}, we show that the 
variation in the eigenfrequencies with respect to the distance between two 
bubbles is more complicated than those expected by Zabolotskaya \cite{2} and 
Doinikov and Zavtrak \cite{3,4}. 
In Sec.~\ref{sec2b}, 
additionally, we discuss theoretically the influence of the mutual interaction 
on the amplitude of the bubble pulsation. 
In Sec.~\ref{sec3}, we show some numerical results of the primary Bjerknes 
force, acting on two interacting bubbles, by using a nonlinear model 
in order to verify the linear theory for the eigenfrequencies. When the 
frequency of an external sound is equivalent to an eigenfrequency of a bubble, 
the force vanishes because of the phase difference of $\pi /2$ between the 
sound and the bubble pulsation. Namely, by observing the points where the force 
vanishes, one can know how the eigenfrequencies changed by the interaction. 

In the present study, an external sound field of weak or moderate amplitude 
with a frequency comparable to the 
partial resonance frequencies of bubbles is assumed.

\section{MATHEMATICAL FORMULATION}
\label{sec1}
A gas bubble levitated in a liquid pulsates when a sound wave is applied. 
The sound pressure at the bubble position drives the pulsation. When other 
bubbles (named ``bubble 2'' $\sim $ ``bubble $N$'', where $N$ is the number of 
existing bubbles) exist near the bubble (bubble 1), the pulsation of bubble 
1 is also driven by the sound wave scattered by the other bubbles. Namely, 
the driving pressure acting on bubble 1, $p_{{\rm d}\,1}$, is expressed as
\begin{equation}
\label{eq1}
p_{{\rm d}\,1} = p_{{\rm ex}}+{\sum\limits_{j = 2}^{N}{p_{{\rm s}\,1j}}},
\end{equation}
where $p_{\rm ex} $ and $p_{{\rm s}\,1j} $ are the sound pressures 
of the external sound field and the scattered wave emitted by bubble $j$, 
respectively, at the position of bubble 1. The pressure of the external 
sound can be considered uniform when the wavelength of the standing wave 
is much larger than the radii of the bubbles and the distance between 
bubbles. By assuming that those bubbles keep being spherically symmetric and 
the surrounding liquid is incompressible, the scattered pressure can be 
estimated with \cite{7}
\begin{equation}
\label{eq2}
p_{{\rm s}\,1j} \approx {\frac{{\rho} }{{r_{1j} }}}
{\frac{{d}}{{dt}}}(R_{j}^{2} \dot {R}_{j} ),
\end{equation}
where $\rho $ is the density of the liquid, $r_{1j} $ is the distance 
between the centers of bubble 1 and bubble $j$, $R_j$ is the radius of 
bubble $j$, and the dot denotes the time derivative.

When the compressibility of the material surrounding the bubbles is completely 
negligible, one can use the RPNNP equation \cite{9}:
\begin{equation}
\label{eq3}
R_1 \ddot R_1  + \frac{3}{2}\dot R_1^2  - \frac{1}{\rho }p_{{\rm w}\,1}
= - \frac{1}{\rho }p_{{\rm d}\,1},
\end{equation}
with
\[
p_{{\rm w}\,1} = \left( P_0  + \frac{{2\sigma }}{{R_{10} }} \right) \left( \frac{{R_{10} }}{{R_1 }} \right)^{3\kappa } - \frac{{2\sigma }}{{R_1 }} - \frac{{4\mu \dot R_1 }}{{R_1 }} - P_0,
\]
for modeling the radial pulsation of bubbles, where $P_0$ is the static 
pressure, $\sigma $ is the surface tension at the bubble surface, $\mu $ is the 
viscosity in the liquid, $\kappa$ is the polytropic exponent of the gas 
inside the bubbles, and $R_{10}$ is the equilibrium radius of bubble 1. 
Substitution of Eqs.~(\ref{eq1}) and (\ref{eq2}) into Eq.~(\ref{eq3}) yields
\[
R_1 \ddot R_1 + \frac{3}{2}\dot R_1^2 - \frac{1}{\rho }p_{{\rm w}\,1} = - \frac{1}{\rho }[p_{{\rm ex}} + \sum\limits_{j = 2}^N {\frac{\rho }{{r_{1j} }}\frac{d}{{dt}}(R_j^2 \dot R_j )} ].
\]
When the number of bubbles is two, this equation is reduced to
\begin{equation}
\label{eq4}
R_1 \ddot R_1 + \frac{3}{2}\dot R_1^2 - \frac{1}{\rho }p_{{\rm w}\,1} = - \frac{1}{\rho }[p_{{\rm ex}}  + \frac{\rho }{D}\frac{d}{{dt}}(R_2^2 \dot R_2 )],
\end{equation}
where $D = r_{12}$ ($ = r_{21}$). 
Exchanging 1 and 2 in the subscripts yields the model equation 
for bubble 2:
\begin{equation}
\label{eq5}
R_2 \ddot R_2 + \frac{3}{2}\dot R_2^2 - \frac{1}{\rho }p_{{\rm w}\,2} = - \frac{1}{\rho }[p_{{\rm ex}} + \frac{\rho }{D}\frac{d}{{dt}}(R_1^2 \dot R_1 )]
\end{equation}
with
\[
p_{{\rm w}\,2}  = \left( P_0  + \frac{{2\sigma }}{{R_{20} }} \right) \left( \frac{{R_{20} }}{{R_2 }} \right)^{3\kappa}  - \frac{{2\sigma }}{{R_2 }} - \frac{{4\mu \dot R_2 }}{{R_2 }} - P_0 .
\]

In the following sections, we present theoretical and numerical discussions 
regarding the above systems of equations (\ref{eq4}) and (\ref{eq5}).

\section{LINEAR ANALYSIS}
\label{sec2}
In this section, some basic properties of Eqs.~(\ref{eq4}) and (\ref{eq5}) are 
discussed by means of linear theory. By assuming that the bubble pulsation 
can be represented as $R_1  = R_{10}  + e_1$, $R_2  = R_{20}  + e_2$, and 
$\left| {e_1 } \right| \ll R_{10}$, $\left| {e_2 } \right| \ll R_{20}$, 
Eqs.~(\ref{eq4}) and (\ref{eq5}) are reduced to the following linear formulae:
\begin{eqnarray}
\label{eq8}
\ddot e_1 + \omega _{10}^2 e_1 + \delta _1 \dot e_1 = - \frac{{p_{{\rm ex}} }}{{\rho R_{10} }} - \frac{{R_{20}^2 }}{{R_{10} D}}\ddot e_2, \\
\label{eq9}
\ddot e_2 + \omega _{20}^2 e_2 + \delta _2 \dot e_2 = - \frac{{p_{{\rm ex}} }}{{\rho R_{20} }} - \frac{{R_{10}^2 }}{{R_{20} D}}\ddot e_1,
\end{eqnarray}
where
\[
\omega _{10} = \sqrt {\frac{1}{{\rho R_{10}^2 }}[3\kappa P_0 + (3\kappa - 1)\frac{{2\sigma }}{{R_{10} }}]}
\]
and
\[
\omega _{20} = \sqrt {\frac{1}{{\rho R_{20}^2 }}[3\kappa P_0 + (3\kappa - 1)\frac{{2\sigma }}{{R_{20} }}]} 
\]
are the partial resonance (angular) frequencies of bubble 1 and bubble 2, 
respectively,
\[
\delta _1 = \frac{{4\mu }}{{\rho R_{10}^2 }} \quad {\rm and} \quad 
\delta _2 = \frac{{4\mu }}{{\rho R_{20}^2 }}.
\]
In the case where $\sigma \approx 0$, the system of equations (\ref{eq8}) and 
(\ref{eq9}) corresponds to that derived by Zabolotskaya \cite{2} on the basis 
of Lagrangian formalism.

\subsection{Eigenfrequencies of the linear system of equations}
\label{sec2a}
Let us analyze in detail the eigenfrequencies of the linear coupled system, 
although a brief discussion is shown in Ref.~\cite{2}. When only one bubble 
exists in an acoustic field and the frequency of the external sound is 
equivalent to the resonance frequency of the bubble, the phase difference 
between the bubble pulsation and the external sound becomes $\pi /2$, as is 
well known. According to this analogy, we call the frequency of an external 
sound an eigenfrequency of the linear coupled system when the phase difference 
between $p_{{\rm ex}}$ and $e_1$ (or $p_{{\rm ex}}$ and $e_2$) becomes 
$\pi /2$. We assume that the external sound pressure at the bubble position is 
written in a form of $p_{{\rm ex}} = - P_a \sin \omega t$. A harmonic 
steady-state solution of the linear coupled system is
\begin{equation}
\label{eq10}
e_1 = K_1 \sin (\omega t + \phi _1 ),
\end{equation}
where
\begin{equation}
\label{eq11}
K_1 = \frac{{P_a }}{{R_{10} \rho }}\sqrt {A_1^2  + B_1^2 },
\end{equation}
\[
\phi _1 = \tan ^{-1} \left( {\frac{{B_1 }}{{A_1 }}} \right),
\]
with
\begin{eqnarray}
\label{eq12}
A_1 = \frac{{H_1 F + M_2 G}}{{F^2 + G^2 }}, \\
\label{eq13}
B_1 = \frac{{M_2 F - H_1 G}}{{F^2 + G^2 }},
\end{eqnarray}
\begin{eqnarray*}
&{\displaystyle F = L_1 L_2  - \frac{{R_{10} R_{20} }}{{D^2 }}\omega ^4  - M_1 M_2,}& \\
&G = L_1 M_2 + L_2 M_1,& \\
&{\displaystyle H_1 = L_2 + \frac{{R_{20} }}{D}\omega ^2,}& \\
&L_1 = (\omega _{10}^2 - \omega ^2), \quad L_2 = (\omega _{20}^2 - \omega ^2),& \\
&M_1 = \delta _1 \omega , \quad M_2 = \delta _2 \omega.&
\end{eqnarray*}
Here the solution for only bubble 1 is shown. The phase difference $\phi _1$ 
becomes $\pi /2$ when
\begin{equation}
\label{eq14}
A_1 = 0.
\end{equation}
It should be noted that a case in which both $A_1$ and $B_1$ become zero does 
not exist. From Eqs.~(\ref{eq12}) and (\ref{eq13}), one obtains
\[
A_1^2 + B_1^2 = \frac{{H_1 ^2  + M_2 ^2 }}{{F^2  + G^2 }}\left( {\frac{{P_a }}{{R_{10} \rho }}} \right)^2 .
\]
The numerator of this equation always has a nonzero value since $M_2 > 0$; this 
result denies the existence of the case where both $A_1 = 0$ and $B_1 = 0$ are 
true. It should be noted also that $F^2  + G^2 $ appearing in the dominator of 
Eq.~(\ref{eq12}) always has a nonzero value. When $G = 0$, $F$ is reduced to
\[
F = - \frac{{M_2 }}{{M_1 }}L_1^2 - \frac{{R_{10} R_{20} }}{{D^2 }}\omega ^4 - M_1 M_2 .
\]
This has a nonzero, negative value because of $M_2 L_1^2 /M_1 \ge 0$, 
$R_{10} R_{20} \omega ^4 /D^2 > 0$ and $M_1 M_2 > 0$. This result means that 
no case exists where both $F = 0$ and $G = 0$ are true. In a consequence, 
Eq.~(\ref{eq14}) is reduced to
\begin{equation}
\label{eq15}
H_1 F + M_2 G = 0 .
\end{equation}
In the following, we analyze this equation.

When the viscous terms in Eq.~(\ref{eq15}) are negligible (but exist), one can 
easily obtain the solutions of this equation. By assuming that 
$M_1 \approx 0$ and $M_2 \approx 0$, one obtains
\[
H_1 F = 0,
\]
or
\begin{equation}
\label{eq16}
F \approx L_1 L_2 - \frac{{R_{10} R_{20} }}{{D^2 }}\omega ^4 = 0
\end{equation}
and
\begin{equation}
\label{eq17}
H_1 = L_2 + \frac{{R_{20} }}{D}\omega ^2 = 0 .
\end{equation}
Equation (\ref{eq16}) corresponds to that given in Ref.~\cite{2}, and is 
symmetric; namely, exchanging 1 and 2 (or 10 and 20) on the subscripts of 
variables in this equation yields the same equation. This means that two 
bubbles have the same eigenfrequencies.

The solutions of Eq.~(\ref{eq16}) are
\begin{equation}
\label{eq18}
\omega _{1 \pm }^2 = \frac{{\omega _{10}^2  + \omega _{20}^2  \pm \sqrt {\left( {\omega _{10}^2 - \omega _{20}^2 } \right)^2  + 4\omega _{10}^2 \omega _{20}^2 \frac{{R_{10} R_{20} }}{{D^2 }}} }}{{2\left( {1 - R_{10} R_{20} / D^2} \right)}}.\end{equation}
Those solutions can be rewritten approximately as follows in the case where the 
distance $D$ is large enough and $\omega _{10} > \omega _{20}$:
\begin{equation}
\label{eq19}
\omega _{1 + }^2  \approx \left[ {1 + \left( {\frac{{\omega _{10}^2 }}{{\omega _{10}^2  - \omega _{20}^2 }}} \right)\left( {\frac{{R_{10} R_{20} }}{{D^2  - R_{10} R_{20} }}} \right)} \right]\omega _{10}^2 
\end{equation}
and
\begin{equation}
\label{eq20}
\omega _{1 - }^2  \approx \left[ {1 - \left( {\frac{{\omega _{20}^2 }}{{\omega _{10}^2  - \omega _{20}^2 }}} \right)\left( {\frac{{R_{10} R_{20} }}{{D^2  - R_{10} R_{20} }}} \right)} \right]\omega _{20}^2 .
\end{equation}
For $D \sim \infty $, the former and the latter one converge to 
$\omega _{10}^2 $ and $\omega _{20}^2$, respectively. As the bubbles approach 
each other, the former increases and the latter decreases; namely, the 
higher eigenfrequency becomes higher and the lower one becomes lower.

In the case of $\omega _{10} < \omega _{20}$ and large $D$, the solutions 
(\ref{eq18}) are reduced to
\begin{equation}
\label{eq21}
\omega _{1 + }^2 \approx \left[ {1 + \left( {\frac{{\omega _{20}^2 }}{{\omega _{20}^2  - \omega _{10}^2 }}} \right)\left( {\frac{{R_{10} R_{20} }}{{D^2  - R_{10} R_{20} }}} \right)} \right]\omega _{20}^2 
\end{equation}
and
\begin{equation}
\label{eq22}
\omega _{1 - }^2  \approx \left[ {1 - \left( {\frac{{\omega _{10}^2 }}{{\omega _{20}^2  - \omega _{10}^2 }}} \right)\left( {\frac{{R_{10} R_{20} }}{{D^2  - R_{10} R_{20} }}} \right)} \right]\omega _{10}^2.
\end{equation}
Same as the result for $\omega _{10} > \omega _{20}$, the higher eigenfrequency 
becomes higher and the lower one becomes lower as $D$ decreases.

The solution of Eq.~(\ref{eq17}) given for the first time is
\begin{equation}
\label{eq23}
\omega _1^2  = \frac{{\omega _{20}^2 }}{{1 - R_{20} /D}}.
\end{equation}
This converges to $\omega _{20}^2$ for an infinite $D$ and increases with 
decreasing $D$. In contrast with Eq.~(\ref{eq16}), Eq.~(\ref{eq17}) is 
asymmetric; namely this serves to break the symmetry of the Zabolotskaya's 
result mentioned above.

The present results show that, when the viscous effect of the surrounding 
liquid is negligible and the radii of the bubbles are not equivalent, the 
bubbles have three asymmetric eigenfrequencies; one is the ``fundamental 
eigenfrequency'' that converges to the partial resonance frequency of a 
corresponding bubble for an infinite $D$, and the remaining two are the 
``subeigenfrequencies'' that converge to the resonance frequency of a 
neighboring bubble. One of the subeigenfrequencies always increases as bubbles 
approach each other. The other subeigenfrequency decreases (increases) and the 
fundamental eigenfrequency increases (decreases) when the partial resonance 
frequency of the bubble is higher (lower) than that of the neighboring bubble.

Now we discuss briefly the case of identical bubbles. When two bubbles 
have the same radii and pulsate in phase, Eq.~(\ref{eq8}) is reduced to
\[
\left( {1 + \frac{{R_{10} }}{D}} \right)\ddot e_1 + \omega _{10}^2 e_1 + \delta _1 \dot e_1 = - \frac{{p_{{\rm ex}} }}{{\rho R_{10} }}.
\]
This equation has only one eigenfrequency \cite{2} of
\begin{equation}
\label{eq24}
\omega _1^2  = \frac{{\omega _{10}^2 }}{{1 + R_{10} /D}} ,
\end{equation}
which converges to $\omega _{10}^2 $ for infinite $D$ and decreases with 
decreasing $D$. Using this example, we here carry out quantitative comparison 
between the model used here and previous ones. In Ref.~\cite{add3}, Kobelev and 
Ostrovskii discussed the eigenfrequency of two interacting bubbles by means of 
a mathematical model in which the surface tension and viscosity are neglected 
and the potential field forming between the bubbles is accurately described by 
the method of images. The model predicts only two eigenfrequencies when 
$R_{10} \ne R_{20}$, as in Zabolotskaya's model, and gives 
$\omega _1 = \sqrt{\ln 2}\,\omega _{10} \simeq 0.832\,\omega _{10}$ 
when $R_{10} = R_{20}$ and both bubbles make contact with each other, i.e., 
$D = 2R_{10}$. In Ref.~\cite{add2}, Zavtrak also discussed the 
eigenfrequency of the present case by means of a mathematical model in which 
multiple scattering between two bubbles is taken into account by the 
Legendre polynomial expansion. In the case of $D = 2R_{10}$, Zavtrak's 
model gives $\omega _1 \simeq 0.832\,\omega _{10}$, which is in close 
agreement with the result obtained by Kobelev and Ostrovskii. However, 
Eq.~(\ref{eq24}) gives $\omega _1 \simeq 0.816\,\omega _{10}$ which is lower 
by about 2 \% 
than these values. This quantitative discrepancy is caused from 
the difference of functions for describing the scattered sound wave by a 
bubble. The amplitude of the scattered wave is assumed to be inversely 
proportional to $D$ in the model used here (See Eq.~(\ref{eq2})), while those 
in the models in Refs.~\cite{add3,add2} 
are approximated by an expansion function. Namely, the amplitude of the 
scattered wave at the position of a neighboring bubble is estimated 
differently. The influence of this difference becomes stronger as $D$ 
decreases. However, all models predict the same result in a qualitative sense, 
which reveals the downward shift of the eigenfrequency.
\begin{figure}
\begin{center}
\epsfig{file=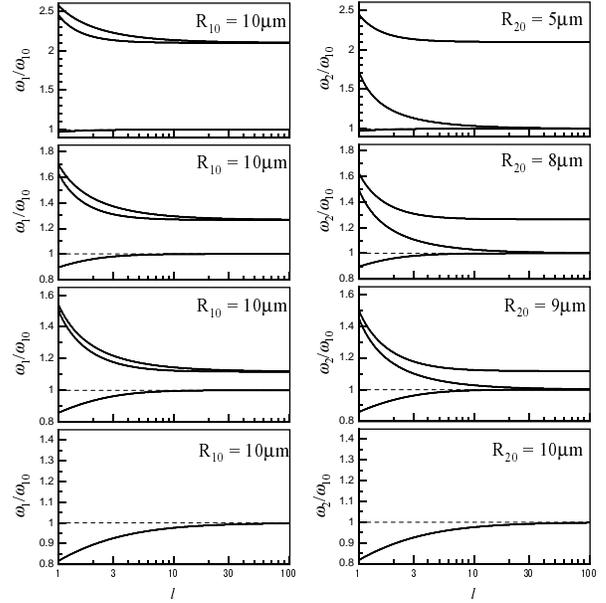,width=8cm}
\end{center}
\caption{The eigenfrequencies of bubbles 1 ($\omega _{1}$) and 2 
($\omega _{2}$) for $R_{0} \sim 10$ $\mu $m and $\mu \approx 0$ normalized by 
$\omega _{10}$. The dashed line denotes the line of 
$\omega _j /\omega _{10} = 1$.}
\label{fig1}
\end{figure}
\begin{figure}
\begin{center}
\epsfig{file=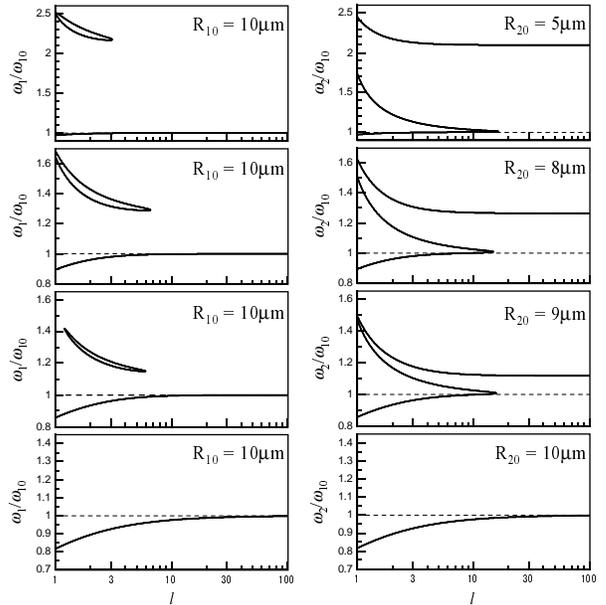,width=8cm}
\end{center}
\caption{The eigenfrequencies for $R_{0} \sim 10$ $\mu $m normalized by 
$\omega _{10}$. The dashed line denotes the line of 
$\omega _j /\omega _{10} = 1$.}
\label{fig2}
\end{figure}

Figure \ref{fig1} shows the eigenfrequencies as a function of 
$l = D/(R_{10} + R_{20})$, in the case where the viscous effect is negligible. 
Here, the parameters are set to $\rho = 1000$ kg/m$^3$, $P_0 = 1$ atm, 
$\sigma = 0.0728$ N/m, $\mu \approx 0$, and $\kappa = 1.4$. Equations 
(\ref{eq18}) and (\ref{eq23}) are used in plotting those graphs since 
Eqs.~(\ref{eq19})$ - $(\ref{eq22}) provide a less-accurate result for small 
$D$. The radius of bubble 1 is fixed to 10 $\mu $m and that of bubble 2 varies 
from 5 $\mu $m to 15 $\mu $m. The eigenfrequencies shown in those graphs 
are normalized with the partial resonance frequency of bubble 1, 
$\omega _{10}$. As discussed above, three eigenfrequencies changing with the 
distance between bubbles appear in each graph except for the case 
of $R_{10} = R_{20}$ where only one decreasing eigenfrequency appears. It 
should be noted that the larger bubble, whose 
partial resonance frequency is lower than that of the smaller bubble, has 
the highest eigenfrequency among those of a pair of bubbles. This result is 
inconsistent with the opinion represented in Refs.~\cite{3,4} in which it is 
assumed that the effective resonance frequency of the smaller bubble rises 
faster than that of the larger bubble when two bubbles approach each other.

\begin{figure}
\begin{center}
\epsfig{file=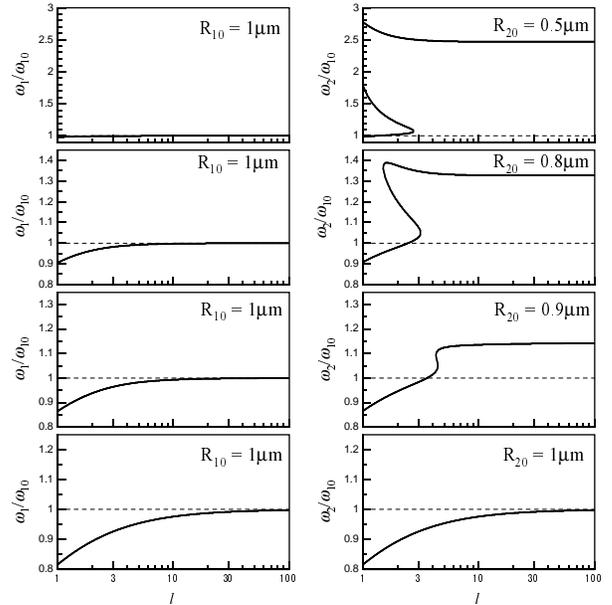,width=8cm}
\end{center}
\caption{The eigenfrequencies for $R_{0} \sim 1$ $\mu $m normalized by 
$\omega _{10}$. The dashed line denotes the line of 
$\omega _j/\omega _{10} = 1$.}
\label{fig3}
\end{figure}

\begin{figure}
\begin{center}
\epsfig{file=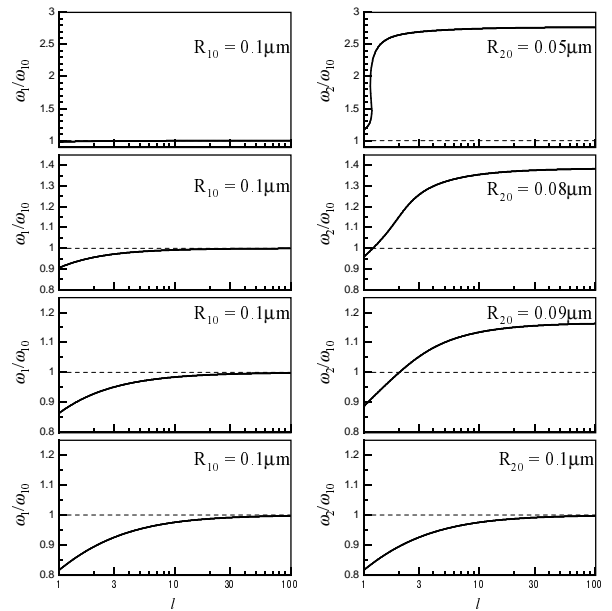,width=8cm}
\end{center}
\caption{The eigenfrequencies for $R_0 \sim 0.1$ $\mu $m normalized by 
$\omega _{10}$. The dashed line denotes the line of 
$\omega _j /\omega _{10} = 1$.}
\label{fig4}
\end{figure}

Now we present numerical solutions of Eq.~(\ref{eq15}) for examining the 
influences of viscosity on eigenfrequency. Figures \ref{fig2}$-$\ref{fig4} show 
the results obtained by using $\mu = 1.137 \times 10^{-3}$ kg/(m s) that 
corresponds to the viscosity of water at room temperature. From those 
figures, we can observe that, as the viscous effect grows strong, i.e., the 
mean radii of the two bubbles become smaller, the subeigenfrequencies vanish 
gradually from the large-distance region (and sometimes from the small-distance 
region) and only the fundamental eigenfrequency remains. In the case of 
$R_1 = 1$ $\mu$m, the subeigenfrequencies of a larger bubble disappear, and 
when the sizes of two bubbles are close each other the fundamental 
eigenfrequency and the higher one of subeigenfrequencies of a smaller bubble 
vanish in the small-distance region.

In the case of $R_1 = 0.1$ $\mu$m, it is difficult to distinguish the 
subeigenfrequencies from the fundamental one since only a smooth curve, which 
decreases monotonically with decreasing $l$, appears in the graphs. (The 
eigenfrequency of the smaller bubble remaining in the small-distance region may 
be the lower one of the subeigenfrequencies.) This result means that in a 
high-viscosity case the effective resonance frequencies of both bubbles 
decrease monotonically as $l$ decreases.

These results show that the influence of the mutual interaction on the 
eigenfrequencies depends on the viscosity of the surrounding material and is 
weakened as the viscous effect becomes stronger, i.e., the threshold of the 
distance for the appearance of the subeigenfrequencies is shortened.

From the results shown in Fig.~\ref{fig2}, we can expect that when 
$R_1 = 10$ $\mu$m, $R_2 = 8$ $\mu$m, and $\omega = 1.4\omega _{10}$, for 
example, the signs of the primary Bjerknes force acting on bubbles 1 and 2 
change twice as the two bubbles approach each other since the 
curves shown in the figure for this case close with the line 
$\omega _j / \omega _{10} = 1.4$ ($j = 1,\;2$) at two points. We can expect 
also from the results shown in Fig.~\ref{fig3} that when $R_1 = 1$ $\mu$m, 
$R_2 = 0.8$ $\mu$m, and $\omega = 1.1\omega _{10}$, the sign of the force 
acting on bubble 1 changes once during the approach since the curves for this 
case close with the line $\omega _1 / \omega _{10} = 1.1$ at one point. Such 
an influence of the mutual interaction on the primary Bjerknes force is 
discussed in Sec.~\ref{sec3}.

\subsection{Amplitude of linear pulsation of two coupled bubbles}
\label{sec2b}
Next, we briefly discuss the amplitude of the bubble pulsation. By 
substituting Eqs.~(\ref{eq12}) and (\ref{eq13}) into Eq.~(\ref{eq11}) 
and rearranging it, one obtains
\begin{equation}
\label{eq25}
K_1^2  = \frac{{H_1 ^2  + M_2 ^2 }}{{F^2  + G^2 }}\left( {\frac{{P_a }}{{R_{10} \rho }}} \right)^2 .
\end{equation}

When $\omega$ is far apart from the partial resonance frequencies of both 
bubbles and the viscous term is negligible, Eq.~(\ref{eq25}) is reduced to
\[
K_1^2  \approx \left( {\frac{{L_2  + R_{20} \omega ^2 /D}}{{L_1 L_2  - R_{10} R_{20} \omega ^4 /D^2 }}} \right)^2 \left( {\frac{{P_a }}{{R_{10} \rho }}} \right)^2 .
\]
Furthermore, neglecting the second order term of $R_{j0} /D$ reduces this to
\begin{eqnarray}
\label{eq26}
K_1^2 && \approx \left( {\frac{{L_2 + R_{20} \omega^2 /D}}{{L_1 L_2}}} \right)^2 \left( {\frac{{P_a }}{{R_{10} \rho }}} \right)^2  \nonumber \\
&& = \left[1 + {\left( {\frac{{R_{20} \omega ^2}}{{\omega _{20}^2 - \omega ^2}}} \right)\frac{1}{D}} \right]^2 \left( {\frac{1}{{L_1 }}} \right)^2 \left( {\frac{{P_a}}{{R_{10} \rho}}} \right )^2 .
\end{eqnarray}
As $D$ decreases, the amplitude $K_1$ increases when $\omega _{20} > \omega$, 
and decreases when $\omega _{20} < \omega $. Physically, this result can be 
understood as follows: When $\omega _{20} $ is sufficiently larger than 
$\omega $, the scattered wave due to bubble 2 oscillates in phase with the 
external sound. In this case, the amplitude of the driving pressure acting on 
bubble 1 increases (See Eq.~(\ref{eq8})); thus, the pulsation amplitude of 
bubble 1 is increased. In contrast, when $\omega _{20}$ is sufficiently smaller 
than $\omega$, the scattered wave due to bubble 2 oscillates out 
of phase with the external sound; this results in the decrease of the pulsation 
amplitude.

Now we consider the case where two bubbles have the same radii and the 
frequency of the external sound is equivalent to their partial resonance 
frequencies, i.e., 
$R_{10} = R_{20} $ and $\omega = \omega _{10} = \omega _{20} $. This setting 
reduces Eq.~(\ref{eq25}) to
\[
K_1^2  = \frac{1}{{(R_{10} ^2 \omega _{10}^2 /D^2  + \delta _1 ^2 )\omega _{10}^2 }}\left( {\frac{{P_a }}{{R_{10} \rho }}} \right)^2 .
\]
This equation shows that the amplitude of the bubble pulsation decreases with 
decreasing $D$. This result can be explained as follows: The decrease of the 
eigenfrequency represented in Eq.~(\ref{eq24}) causes the difference between 
the sound frequency and the effective resonance frequency; resultantly, the 
amplitude of pulsation decreases. Such a decrease in the pulsation amplitude of 
identical two small near-resonance bubbles ($R_0 = 1.5$ $\mu$m) in a viscous 
liquid is shown in Ref.~\cite{11} by employing a direct-numerical-simulation 
technique. The present theory may explain the numerical result.

\section{NUMERICAL EXPERIMENT}
\label{sec3}
In this section, we show numerical results obtained by using the full 
system of nonlinear equations [(\ref{eq4}) and (\ref{eq5})] and sufficiently 
small amplitude of sound in order to verify the linear theory given in the last 
section. We investigate the variations in eigenfrequencies by observing the 
primary Bjerknes force which is a time-averaged force acting on a bubble when 
the external sound field has a spatial gradient. The primary Bjerknes force 
\cite{12} is expressed as
\begin{equation}
\label{eq27}
{\bf F}_j = - \frac{{4\pi }}{3}\left\langle {R_j^3 \; \nabla p_{{\rm ex}} } \right\rangle,
\end{equation}
where $\left\langle \cdots \right\rangle$ denotes the time average over a 
period of the external sound. As was mentioned previously, the phase difference 
between the external sound and the bubble pulsation becomes $\pi /2$ when the 
sound frequency is equivalent to the resonance frequency of a bubble. In this 
situation, furthermore, the primary Bjerknes force vanishes as the result of 
the inner product by Eq.~(\ref{eq27}). This means that we can observe the 
eigenfrequencies of interacting bubbles by examining the conditions under which 
the magnitude of the force becomes zero.

Here we assume that the external standing wave field is in a form of
\[
p_{{\rm ex}} (z,t) = - P_a (z)\sin (\omega t)
\]
with $P_a (z) = P_s \sin (kz)$, where $z$ is the spatial coordinate vertical to 
the pressure nodes, $P_a (z)$ is the amplitude of the standing wave field at 
$z$, and $k$ is the wave number of the standing wave field. This assumption 
yields
\begin{equation}
\label{eq28}
\nabla p_{{\rm ex}} (z,t) = k\frac{{\cos (kz)}}{{\sin (kz)}}p_{{\rm ex}} (z,t) \,{\bf z} ,
\end{equation}
where ${\bf z}$ is the unit vector in the $z$ direction. Substitution of 
Eq.~(\ref{eq28}) into Eq.~(\ref{eq27}) results in
\[
{\bf F}_j = - \frac{{4\pi }}{3}k\frac{{\cos (kz_j )}}{{\sin (kz_j )}}f_j {\bf z}
\]
with
\begin{equation}
\label{eq29}
f_j = \left\langle {R_j^3 \; p_{{\rm ex}} (z_j ,t)} \right\rangle ,
\end{equation}
where $z_j$ is the position of bubble $j$ in the $z$ coordinate. The method for 
computing the force is the same as what used in Refs.~\cite{7,8}. The full 
system of nonlinear equations of (\ref{eq4}) and (\ref{eq5}) is solved 
numerically by the 4th-order Runge-Kutta 
method, and $f_j$ is then computed with $R_{j}$ given by the numerical 
computation. The time average in Eq.~(\ref{eq29}) is done over a sufficiently 
large period after the transition has decayed. Here we assume that the 
wavelength of the external sound is much larger than the distance between the 
bubbles, and thus, $p_{{\rm ex}} (z_1,t) = p_{{\rm ex}} (z_2,t)$.

\end{multicols}

\begin{figure}
\begin{center}
\epsfig{file=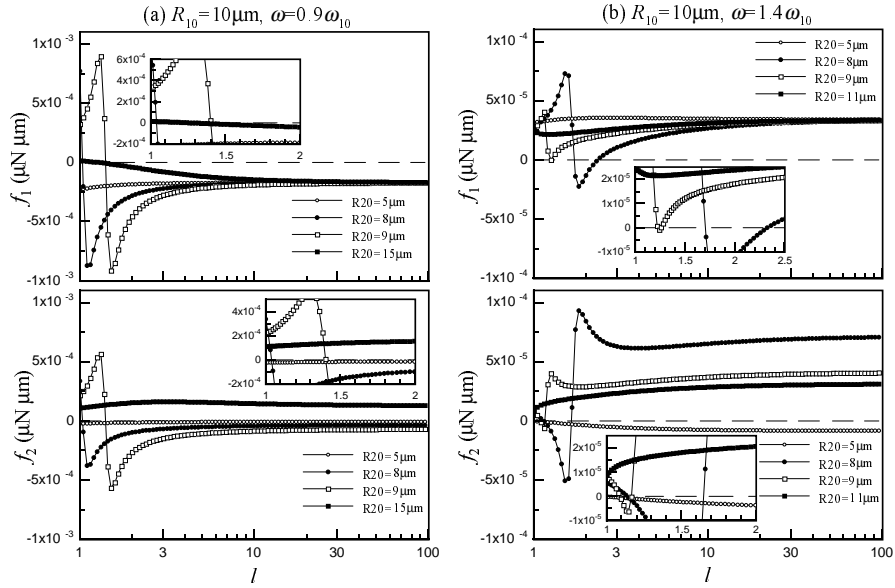,width=12cm}
\end{center}
\caption{$f_j - l$ curves for $R_{10} = 10$ $\mu $m and different values of 
$R_{20} $. The sound frequency is set to $\omega  = 0.9\omega _{10}$ (a) and 
$1.4\omega _{10}$ (b), and the sound amplitude is $0.001P_0$. 
The theoretical values of ($l_1 $, $l_2 $), where $l_1 $ and $l_2$ are the 
normalized distances at which $f_1 = 0$ and $f_2 = 0$, respectively, 
given by Eq.~(\ref{eq15}) are (a): (1.038, 1.038) for $R_{20} = 8$ $\mu $m, 
(1.402, 1.402) for $R_{20} = 9$ $\mu $m, 
(1.279, None) for $R_{20} = 15$ $\mu $m and (b): (None, 1.362) for 
$R_{20} = 5$ $\mu $m, (1.703 and 2.330, 1.139 and 1.649) for 
$R_{20} = 8$ $\mu $m, (1.224 and 1.258, 1.089 and 1.166) for 
$R_{20} = 9$ $\mu $m.}
\label{fig5}
\end{figure}
\begin{figure}
\begin{center}
\epsfig{file=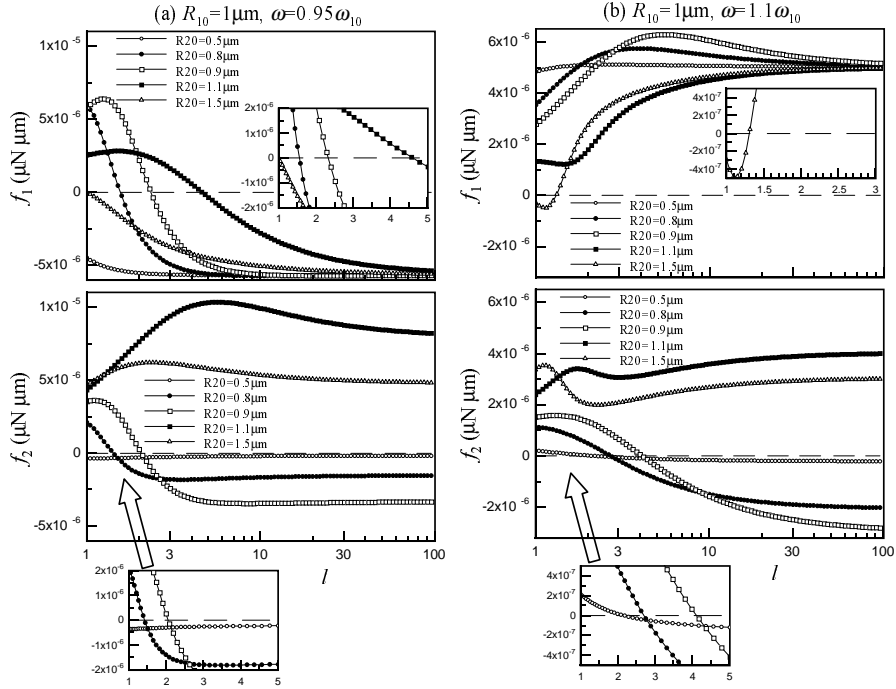,width=12cm}
\end{center}
\caption{$f_j - l$ curves for $R_{10} = 1$ $\mu $m and different values of 
$R_{20} $. The sound frequency is set to $\omega = 0.95\omega _{10}$ (a) 
and $1.1\omega _{10}$ (b), and the sound amplitude is $0.01P_0 $. The 
theoretical values of ($l_1 $, $l_2 $) given by Eq.~(\ref{eq15}) are (a): 
(1.574, 1.427) for $R_{20} = 0.8$ $\mu $m, (2.355, 2.077) for 
$R_{20} = 0.9$ $\mu $m, (4.761, None) for $R_{20} = 1.1$ $\mu $m, 
(1.049, None) for $R_{20} = 1.5$ $\mu $m and (b): (None, 2.637) for 
$R_{20} = 0.5$ $\mu $m, (None, 2.789) for $R_{20}$ = 0.8 $\mu $m, 
(None, 4.279) for $R_{20} = 0.9$ $\mu $m, (1.309, None) for $R_{20} = 1.5$ 
$\mu $m.}
\label{fig6}
\end{figure}

\begin{multicols}{2}

\begin{figure}
\begin{center}
\epsfig{file=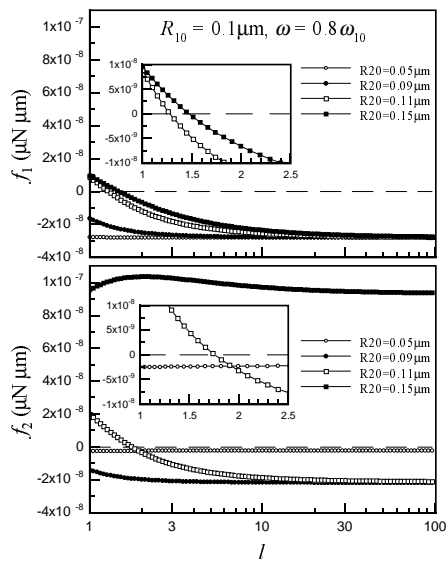,width=6cm}
\end{center}
\caption{$f_j - l$ curves for $R_{10}  = 0.1$ $\mu $m and different values 
of $R_{20} $. The sound frequency is $\omega = 0.8\omega _{10} $ and the 
sound amplitude is $0.1P_0 $. The theoretical values of ($l_1 $, $l_2 $) 
given by Eq.~(\ref{eq15}) are (1.279, 1.803) for $R_{20} = 0.11$ $\mu $m, 
(1.474, None) for $R_{20} = 0.15$ $\mu $m.}
\label{fig7}
\end{figure}

Figures \ref{fig5}$-$\ref{fig7} show $f_j-l$ curves for the cases of 
$R_{10} = 10$ $\mu$m, $R_{10} = 1$ $\mu$m and $R_{10}  = 0.1$ $\mu$m, 
respectively. Here the system of Eqs.~(\ref{eq4}) and (\ref{eq5}) is used; 
namely, the compressibility of the surrounding material is neglected. The 
amplitude of the external sound at the bubble position is assumed as 
$P_a = 0.001P_0$, $0.01P_0$ and $0.1P_0$ for $R_{10} = 10$ $\mu$m, $1$ $\mu$m 
and $0.1$ $\mu$m cases, respectively. The frequency $\omega$ of the external 
sound is set to $0.9\omega _{10} $ and $1.4\omega _{10}$ for $R_{10} = 10$ 
$\mu$m, $0.95\omega _{10}$ and $1.1\omega _{10}$ for $R_{10} = 1$ $\mu$m, and 
$0.8\omega _{10}$ for $R_{10} = 0.1$ $\mu$m. In those graphs, we can observe 
some points where the primary Bjerknes force vanishes. The normalized distances 
at the points, given theoretically by solving Eq.~(\ref{eq15}) as a function of 
$D$ with the given $\omega$, are listed in the captions of those 
figures. These theoretical values are in good agreement with the numerical ones 
represented in those graphs. This result confirms our prediction that the sign 
of the primary Bjerknes force changes at distances where one of the 
eigenfrequencies becomes equivalent to the sound frequency.

\section{CONCLUSION}
In the present study, the dependency of the eigenfrequencies of two mutually 
interacting bubbles on the distance between the bubbles was theoretically 
determined concretely and in detail. The dependency is more intricate than 
those predicted or assumed previously by several authors \cite{2,3,4}. It was 
shown also that the eigenfrequencies depend also on the viscosity of the 
bubbles' surrounding material, unlike the partial natural frequency. Moreover, 
the present theory was verified 
numerically by observing the primary Bjerknes force acting on bubbles. These 
present results may be reasonable for understanding the reversal of sign 
of the secondary Bjerknes force \cite{2,3,4,7} and the pattern formations of 
bubbles or clusters of them in an acoustic field \cite{1,5}. As was pointed 
out in Sec.~\ref{sec2a}, Zabolotskaya's theory predicts that both bubbles 
interacting with each other have the same eigenfrequencies. This result may 
give rise to simultaneous phase changes of the bubbles, e.g., from in-phase to 
out-of-phase with the external sound, and, as a result, may deny the reversal 
of the sign. The asymmetric eigenfrequencies derived in the present paper may 
be able to describe the reversal of the sign more accurately. (This subject 
will be discussed in a future paper.) Furthermore, these present results will 
also affect understandings on some related subjects such as acoustic 
localization \cite{13} and superresonances \cite{14}, which were investigated 
by employing a system containing only identical bubbles.

\end{multicols}

\end{document}